\documentclass[aps,prd,onecolumn,notitlepage,12pt,a4paper]{revtex4-1}

\usepackage[utf8]{inputenc}
\usepackage{amsmath}
\usepackage{amsfonts}
\usepackage{amssymb}
\usepackage{amsthm}
\usepackage{graphicx}
\usepackage{bm}
\usepackage{hyperref}

\allowdisplaybreaks

\begin{document}

\title{An effective field theory approach to hybrids}
\author{Matthias Berwein}
\affiliation{Technische Universität München, Physik-Department T30f, James-Franck-Str.\ 1, 85747 Garching, Germany}
\email{matthias.berwein@mytum.de}

\begin{abstract}
 Heavy quarkonium hybrids are studied in an effective field theory framework. Coupled and uncoupled Schrödinger equations are obtained for different quantum numbers of the hybrid states. The results are discussed and compared to other approaches.
\end{abstract}

\maketitle

We summarize here the methods and results of our more detailed paper on hybrids published in~\cite{Berwein:2015vca}. Heavy quarkonia are bound states of a quark and an antiquark of charm or bottom flavor, which are classified by their angular momentum and spin quantum numbers $J^{PC}$ as well as their radial excitation. However, QCD also allows for bound states with gluonic or light quark excitations, which are called hybrids or tetraquarks respectively. Such excitations are believed to account for the exotic states observed in experiment, collectively called $X$, $Y$, or $Z$. They allow for $J^{PC}$ quantum numbers which cannot be obtained in a pure quark-antiquark system.

The great advantage in studying systems with heavy quarks is that usually their mass $M$ is larger than any other energy scale in the system. This means that even when the coupling constant $g$ is large and cannot be used as a perturbative parameter, still one can perform an expansion in $1/M$. A systematic way of achieving this is through effective field theories (EFTs), where the expansion is performed at the Lagrangian level. The EFT obtained from QCD by integrating out the heavy quark mass is called nonrelativistic QCD (NRQCD)~\cite{Caswell:1985ui,Bodwin:1994jh}. The effective degrees of freedom are two Pauli spinor fields $\psi$ and $\chi$, which annihilate a heavy quark and create a heavy antiquark respectively, so-called soft gluons, i.e., with momenta much smaller than $M$, as well as $n_f$ flavors of light quarks $q$. The Hamiltonian of NRQCD is also given as an expansion in $1/M$:
\begin{align}
 &\mathcal{H}_\mathrm{NRQCD}=\mathcal{H}^{(0)}+\mathcal{H}^{(1)}+\mathcal{O}\left(M^{-2}\right)\,,\\
 &\mathcal{H}^{(0)}=\frac{1}{2}\int d^3x\left(\bm{E}^a\cdot\bm{E}^a+\bm{B}^a\cdot\bm{B}^a\right)-\sum_{j=1}^{n_f}\int d^3x\,\bar{q}_ji\bm{D}\cdot\bm{\gamma}q_j\,,\\
 &\mathcal{H}^{(1)}=-\frac{1}{2M}\int d^3x\,\psi^\dagger\left(\bm{D}^2+c_Fg\bm{B}\cdot\bm{\sigma}\right)\psi+\frac{1}{2M}\int d^3x\,\chi^\dagger\left(\bm{D}^2+c_Fg\bm{B}\cdot\bm{\sigma}\right)\chi\,.
\end{align}
The physical states are required to satisfy the Gauss law
\begin{equation}
(\bm{D}\cdot\bm{E})^a\,|\mathrm{phys}\rangle=g\Bigl(\psi^\dagger T^a\psi+\chi^\dagger T^a\chi+\sum_{j=1}^{n_f}\bar{q}_j\gamma^0T^aq_j\Bigr)|\mathrm{phys}\rangle\,.
\label{gausslaw}
\end{equation}
In the following we will neglect the light quarks, i.e., $n_f=0$.

The principles of this $1/M$ expansion in physical observables are the same as in perturbation theory of ordinary quantum mechanics. We have a Hamiltonian that is split into a main part $\mathcal{H}^{(0)}$ and a perturbation $\mathcal{H}^{(1)}$. In order to find the spectrum of the full Hamiltonian, we first determine the spectrum of the main part and then calculate corrections from the perturbation. The main part $\mathcal{H}^{(0)}$ corresponds to the theory in the limit of infinite quark mass, also called the static limit. In this limit, the heavy quark and antiquark act only as static color sources, and accordingly they do not appear in the Hamiltonian (although they are still connected to the gluons through the Gauss law). So the positions $\bm{x}_1$ and $\bm{x}_2$ for the heavy quark and the antiquark are good quantum numbers in this limit. We can write the spectrum as
\begin{align}
 \mathcal{H}^{(0)}|\underline{n};\bm{x}_1,\bm{x}_2\rangle^{(0)}&=E_n^{(0)}(r)|\underline{n};\bm{x}_1,\bm{x}_2\rangle^{(0)}\,,\\
 \mathrm{with}\hspace{20pt}^{(0)}\langle\underline{n}';\bm{x}_1',\bm{x}_2'|\underline{n};\bm{x}_1,\bm{x}_2\rangle^{(0)}&=\delta_{n'n}\delta(\bm{x}_1'-\bm{x}_1)\delta(\bm{x}_2'-\bm{x}_2)\,,
\end{align}
where $n$ stands for the set of all other quantum numbers, and the static energies $E_n^{(0)}$ depend only on the relativ quark-antiquark distance $r=|\bm{x}_1-\bm{x}_2|$ because of translational and rotational invariance.

In general, the spectrum of $\mathcal{H}^{(0)}$ cannot be obtained analytically, because it would correspond to the solution of QCD without heavy quarks, or Yang-Mills theory for $n_f=0$. However, at least the static energies can be obtained numerically in the following way: consider some arbitrary state $|X_n\rangle$ in the static theory that has the same quantum numbers as $|\underline{n};\bm{x}_1,\bm{x}_2\rangle^{(0)}$; the large time $T$ transition amplitude of this state is given by
\begin{equation}
 \langle X_n(T)|X_n(0)\rangle=\langle X_n|e^{-i\mathcal{H}^{(0)}T}|X_n\rangle=\sum_n\int d^3x_1d^3x_2\,\left|\langle X_n|\underline{n};\bm{x}_1,\bm{x}_2\rangle^{(0)}\right|^2e^{-iE_n^{(0)}(r)T}\,.
\end{equation}
We have inserted the unity operator $\sum_n\int d^3x_1d^3x_2\,|\underline{n};\bm{x}_1,\bm{x}_2\rangle^{(0)}{}^{(0)}\langle\underline{n};\bm{x}_1,\bm{x}_2|$, where the sum contains in principle all quantum numbers $n$, however, in this equation the sum contains only excited states with the same quantum numbers as $|X_n\rangle$, because the overlap of $|X_n\rangle$ with any other eigenstate is zero. In the large time limit, this correlator is then dominated by the lowest static energy with these quantum numbers, so
\begin{equation}
 E_n^{(0)}(r)=\lim_{T\to\infty}\frac{i}{T}\ln\langle X_n(T)|X_n(0)\rangle\,,
\end{equation}
where some analytic continuation to complex times $T$ is implied. By extending the set of states $|X_n\rangle$, also excited static energies with quantum numbers $n$ can be projected out.

In this way, the lowest static energies have been determined on the lattice in the quenched approximation in~\cite{Juge:2002br} and~\cite{Bali:2003jq}, whose data we have used in~\cite{Berwein:2015vca}. The different static energies are classified by their quantum numbers, which represent the transformation properties under the symmetries of the static system. The symmetry group for two static particles of opposite charge is the same as for a cylinder and called $D_{\infty\,h}$, where the parity transformation $P$ has to be extended by charge conjugation $C$. The elementary transformations of this group are rotations around the quark-antiquark axis, $CP$ transformation, and reflections across a plane that contains the axis. The corresponding quantum numbers are labeled as $\Lambda_\eta^\sigma$, where $\Lambda$ is the rotational quantum number and can be interpreted as the absolute value of the projection of the gluonic angular momentum operator on the quark-antiquark axis, the sign $\eta$ is the eigenvalue under $CP$ transformations, and $\sigma$ is the sign under reflections. $\Lambda$ can take integer values $0,1,2,\dots$ for which traditionally capital Greek letters $\Sigma,\Pi,\Delta,\dots$ are used, for $\eta$ the labels $g$ for plus and $u$ for minus are used, and the quantum number $\sigma$ appears only for $\Lambda=0$, because $\Lambda\geq1$ representations have two degenerate components with different values of $\sigma$, so $\sigma$ is irrelevant for the static energies. According to the lattice results, the ground state has $\Sigma_g^+$ quantum numbers, which corresponds to standard quarkonium, and the lowest gluonic excitations are $\Pi_u$ and $\Sigma_u^-$ states.

Complementary information can be obtained in the short quark-antiquark distance limit $r\to0$. In this limit another EFT called potential NRQCD (pNRQCD)~\cite{Pineda:1997bj,Brambilla:1999xf} can be obtained by integrating out the scale $1/r$; the resulting expansion in $r$ has the form of a multipole expansion. After integrating out $1/r$, the heavy quark and antiquark fields can no longer be resolved individually, so the effective degrees of freedom are quarkonium fields, which can appear in either a color singlet or octet configuration, called $S$ or $O^a$ respectively, as well as so-called ultrasoft gluons with momenta smaller than $1/r$. The Hamiltonian of pNRQCD will be labeled with a double index, the first referring to the power of $1/M$ and the second to the power of $r$. The first terms in the multipole expansion of the static Hamiltonian are given by
\begin{align}
 \mathcal{H}^{(0,0)}&=\frac{1}{2}\int d^3R\left(\bm{E}^a\cdot\bm{E}^a+\bm{B}^a\cdot\bm{B}^a\right)+\int d^3Rd^3r\left(S^\dagger V_s(r)S+O^{a\,\dagger}V_o(r)O^a\right)\,,\\
 \mathcal{H}^{(0,1)}&=-\int d^3Rd^3r\left[\frac{V_A(r)}{\sqrt{6}}\left(O^{a\,\dagger}\bm{r}\cdot g\bm{E}^aS+S^\dagger\bm{r}\cdot g\bm{E}^aO^a\right)+\frac{V_B(r)}{2}d^{abc}O^{a\,\dagger}\bm{r}\cdot g\bm{E}^bO^c\right]\,,
\end{align}
where $d^{abc}$ is the symmetric structure constant, $\bm{R}$ and $\bm{r}$ are the center-of-mass coordinate and the relative distance respectively, and the quarkonium fields $S$ and $O^a$ depend on both coordinates, while the gluons depend only on $\bm{R}$.

In this theory it is fairly straightforward to write down suitable $|X_n\rangle$ states to determine the static energies. For the first two gluonic excitations we use
\begin{equation}
 |X_{\Pi_u}\rangle=\bm{r}\times g\bm{B}^aO^{a\,\dagger}|\mathrm{vac}\rangle\hspace{20pt}\mathrm{and}\hspace{20pt}|X_{\Sigma_u^-}\rangle=\bm{r}\cdot g\bm{B}^aO^{a\,\dagger}|\mathrm{vac}\rangle\,.
\end{equation}
We see that the $\Pi_u$ state has two independent components, while $\Sigma_u^-$ has only one. The static energies we get from these states are identical at leading order:
\begin{align}
 E_{\Pi_u}^{(0)}(r)&=\lim_{T\to\infty}\frac{i}{T}\ln\langle X_{\Pi_u}(T)|X_{\Pi_u}(0)\rangle=V_o(r)+\Lambda_B+\mathcal{O}\left(r^2\right)\,,\\
 E_{\Sigma_u^-}^{(0)}(r)&=\lim_{T\to\infty}\frac{i}{T}\ln\langle X_{\Sigma_u^-}(T)|X_{\Sigma_u^-}(0)\rangle=V_o(r)+\Lambda_B+\mathcal{O}\left(r^2\right)\,,\\
 \mathrm{with}\hspace{20pt}\Lambda_B&=\lim_{T\to\infty}\frac{i}{T}\ln\langle\mathrm{vac}|B_i^{a}(T)\phi^{ab}(T,0)B_i^{b}(0)|\mathrm{vac}\rangle\,,
\end{align}
where $\phi(T,0)$ is a straight Wilson line in the adjoint representation from time $0$ to $T$ ensuring the gauge invariance of the expression. At $\mathcal{O}\left(r^2\right)$, interactions from $\mathcal{H}^{(0,1)}$ come into play, which introduce a splitting between the two static energies, but the calculation of these corrections requires the nonperturbative evaluation of correlators of four gauge fields, which are currently not available. Already for the so-called magnetic gluelump mass $\Lambda_B$ one has to rely on nonperturbative determinations, as performed in~\cite{Bali:2003jq}. So we have determined those quadratic terms instead from a fit to the lattice data of the static energies from~\cite{Juge:2002br} and~\cite{Bali:2003jq}.

But what this leading pNRQCD result shows clearly is the approximate degeneracy of the two static energies for small distances, which becomes exact in the $r\to0$ limit. This requires some care when calculating perturbative corrections to the static result. In naive perturbation theory for a generic theory with Hamiltonian $H^{(0)}$ and spectrum $|n\rangle^{(0)}$, the first two corrections to the energy due to the perturbation $H^{(1)}$ are given through
\begin{equation}
 E_n=E_n^{(0)}+{}^{(0)}\langle n|H^{(1)}|n\rangle^{(0)}+\sum_{m\neq n}\frac{\,\left|{}^{(0)}\langle n|H^{(1)}|m\rangle^{(0)}\right|^2}{E_n^{(0)}-E_m^{(0)}}+\dots\,.
\end{equation}
However, if two energies are nearly degenerate, as in our case the short distance static energies of $\Pi_u$ and $\Sigma_u^-$, then the second correction term cannot be considered small compared to the first and the perturbative series does not converge. Instead, one has to find the states $|h\rangle$ in the space spanned by the nearly degenerate states $|n\rangle^{(0)}$ which diagonalize the full Hamiltonian $H^{(0)}+H^{(1)}$ on that space:
\begin{equation}
 \langle h'|H^{(0)}+H^{(1)}|h\rangle=E_h^{(1)}\delta_{h'h}\hspace{20pt}\mathrm{with}\hspace{20pt}|h\rangle\in\mathrm{span}\left\{|n\rangle^{(0)}\left|E_n^{(0)}\approx E_{n'}^{(0)}\,\forall n,n'\right.\right\}\,.
\end{equation}
The next correction to the energy is then given by
\begin{equation}
 E_h=E_h^{(1)}+\sum_{m\notin\{h\}}\frac{\,\left|\langle h|H^{(1)}|m\rangle^{(0)}\right|^2}{E_h^{(1)}-E_m^{(0)}}+\dots\,,
\end{equation}
and since the sum now contains no nearly degenerate states the denominator does not become large.

In our case there are in fact two sources for degeneracy: on the one hand there is the short distance behavior of the $\Pi_u$ and $\Sigma_u^-$ static energies, on the other hand we see that $\mathcal{H}^{(1)}$ contains derivatives, which connect states at an infinitesimal difference in their coordinates, and thus the corresponding static energies are naturally also degenerate. The last point can be readily understood in that $\mathcal{H}^{(1)}$ goes beyond the static limit, and for nonstatic quarks the positions are no longer good quantum numbers. Consequently, the lowest hybrid states at leading order beyond the static limit will be a linear combination of the static $\Pi_u$ and $\Sigma_u^-$ states, each equipped with a wave function.

Since the degeneracy between $\Pi_u$ and $\Sigma_u^-$ exists only for small distances $r$, we will restrict the calculation of the first nonstatic corrections to the leading order in $r$. The corresponding Hamiltonian in pNRQCD is given by
\begin{equation}
 \mathcal{H}^{(1,-2)}=-\frac{1}{M}\int d^3Rd^3r\left(S^\dagger\bm{\nabla}_r^2S+O^{a\,\dagger}\bm{\nabla}_r^2O^a\right)\,.
\end{equation}
So at this order $\bm{R}$ in fact remains a good quantum number and the wave functions have to depend only on $\bm{r}$. In order to calculate the corrections, we need the actual static states, not just $|X_n\rangle$ states that have a nonvanishing overlap. Even though those are unknown, we can introduce a gluonic operator $\bm{G}_B$ for them and write
\begin{equation}
 |\underline{n};\bm{x}_1,\bm{x}_2\rangle=\bm{\hat{n}}\cdot \bm{G}_B^aO^{a\,\dagger}|\mathrm{vac}\rangle+\mathcal{O}(r)\,,\hspace{20pt}\mathrm{with}\hspace{20pt}\langle\mathrm{vac}|(G_B)_i^a(G_B)_j^b|\mathrm{vac}\rangle=\frac{\delta_{ij}\delta^{ab}}{8}\,,
\end{equation}
where $\bm{\hat{n}}$ are some unit projection vectors perpendicular or parallel to the quark-antiquark axis for $\Pi_u$ or $\Sigma_u^-$ respectively. At leading order, the details of the operator $\bm{G}_B$ are irrelevant, the only important property is the orthogonality relation given above.

The matrix elements of $\mathcal{H}^{(0)}+\mathcal{H}^{(1)}$ between these hybrid states
\begin{equation}
 |\mathrm{Hyb}\rangle=\sum_n\int d^3r\,\bm{\hat{n}}\cdot \bm{G}_B^aO^{a\,\dagger}|\mathrm{vac}\rangle\psi_n(\bm{r})+\mathcal{O}(r)
\end{equation}
are then given by
\begin{equation}
 \langle\mathrm{Hyb}'|\mathcal{H}^{(0)}+\mathcal{H}^{(1)}|\mathrm{Hyb}\rangle=\sum_{n'n}\int d^3r\,\psi^{\prime\,*}_{n'}(\bm{r})\left(E_n^{(0)}(r)\delta_{n'n}-\hat{n}'_i\frac{\nabla_r^2}{M}\hat{n}_i\right)\psi_n(\bm{r})\,.
\end{equation}
The wave functions therefore have to be eigenfunctions of a matrix-valued differential operator, i.e., we have a coupled Schrödinger equation, where the coupling comes from the action of $\bm{\nabla}_r^2$ on the projection vectors $\bm{\hat{n}}$ and $\bm{\hat{n}}'$. Like in ordinary Schrödinger equations, the wave functions can be split into an orbital and a radial part. The orbital wave functions satisfy a modified version of the defining differential equation of the spherical harmonics and are labeled by quantum numbers $l$ and $m$, which correspond to the eigenvalues of the combined angular momentum operator of the gluons and the relative quark-antiquark motion. The radial Schrödinger equation then is found to decouple into two parts: the first part is still coupled but now has only two components:
\begin{equation}
 \left[-\frac{1}{Mr^2}\,\partial_rr^2\partial_r+\frac{1}{Mr^2}\begin{pmatrix} l(l+1)+2 & 2\sqrt{l(l+1)} \\ 2\sqrt{l(l+1)} & l(l+1) \end{pmatrix}+\begin{pmatrix} E_\Sigma^{(0)} & 0 \\ 0 & E_\Pi^{(0)} \end{pmatrix}\right]\begin{pmatrix} \psi_\Sigma \\ \psi_\Pi \end{pmatrix}=\mathcal{E}\begin{pmatrix} \psi_\Sigma \\ \psi_\Pi \end{pmatrix}\,,
\end{equation}
while the second part is an uncoupled Schrödinger equation:
\begin{equation}
 \left[-\frac{1}{Mr^2}\,\partial_r\,r^2\,\partial_r+\frac{l(l+1)}{Mr^2}+E_\Pi^{(0)}\right]\psi_\Pi=\mathcal{E}\,\psi_\Pi\,.
\end{equation}
The two Schrödinger equations correspond to opposite parity solutions; also note that the first equation itself decouples for $l=0$ and only the solution for $\psi_\Sigma$ is physical. The energy eigenvalue $\mathcal{E}$ then gives the mass of the hybrid as $m_\mathrm{Hyb}=2M+\mathcal{E}$.

In~\cite{Berwein:2015vca} we have compared the results of these Schrödinger equations using the potentials obtained from pNRQCD and the lattice data from~\cite{Juge:2002br} and~\cite{Bali:2003jq} with various other investigations of hybrids. There are several experimental candidates in the mass range we obtain, although at this point it is not possible to make a clear identification. Another calculation of hybrid masses has been done in the framework of the Born-Oppenheimer approximation in~\cite{Braaten:2014qka}. While similar in approach and leading to comparable results, they do not obtain the splitting between opposite parity states that emerges clearly in the EFT treatment. There has also been a direct lattice computation of exotic charmonium states in~\cite{Liu:2012ze}. The masses of the hybrid candidates identified in this paper have to be averaged over the different spin configurations for a comparison with our results; there seems to be an overall shift, but in particular the mass differences between different $J^{PC}$ multiplets agree well with ours. For a more detailed discussion and graphical representation of our results, the interested reader is referred to~\cite{Berwein:2015vca}.

\acknowledgments

We thank Colin Morningstar and Gunnar Bali for giving us access to their lattice data for the static energies. We acknowledge support from the DFG cluster of excellence ``Origin and structure of the universe'' (www.universe-cluster.de).

\bibliography{hyb}
\bibliographystyle{apsrev4-1}

\end{document}